\documentclass[twocolumn,aps,showpacs,groupedaddress,floatfix,pre]{revtex4-1}
\usepackage{amsmath}% Incluse math package
\usepackage{graphicx}% Include figure files
\usepackage{dcolumn}% Align table columns on decimal point
\usepackage{bm}% bold math
\usepackage{verbatim}
\usepackage{color}
\usepackage{hyperref}
\usepackage{ulem}
\usepackage{float}%to control floating points
\usepackage{titlesec}
\titlespacing*{\subsection}{0pt}{1.1\baselineskip}{\baselineskip}
\begin{document}
\setlength{\parskip}{0mm}
\title{Morphology of cooperatively rearranging regions in active glass formers}
\author{Dipanwita Ghoshal}
\email{dipanwita.ghoshal.28@gmail.com}
\affiliation{Department of Physics, Indian Institute of Technology Madras, Chennai, Tamil Nadu 600036, India}

\author{Ashwin Joy}

\affiliation{Department of Physics, Indian Institute of Technology Madras, Chennai, Tamil Nadu 600036, India}
\date{\today}

\begin{abstract}
Super cooled liquids display increasingly heterogeneous dynamics as temperature is lowered towards the glass transition ($T_{g}$). A hallmark of this dynamical heterogeneity is the spontaneous emergence of cooperative rearranging regions (CRRs) composed of fast moving particles. While these CRRs in passive glass formers have been explored in great detail, thus understanding is severely limited in active glass formers. The existing consensus on the morphology of CRRs in a passive glass former prioritizes its fast subsets, composed of fast moving particles. In the present study, we focus on a synthetic athermal active glass former and show an equal contribution for the morphology of CRRs from  slow subsets as well. Both these subsets exhibit an exponential  distribution in their structure which strongly correlates with the existence of CRRs. Interestingly, we also observe that the fractal dimensions ($d_{\text{f}}$) of these subsets share both string and compact like morphology that tends to vary in opposite fashion with the control parameters, namely the persistent time ($\tau_{p}$) and the effective temperature ($T_{\text{eff}}$). The fractal dimension $d_{\text{f}}$ measures the roughness or put simply the compactness of fractal objects at their boundaries. For a compact object, the molecules  at boundary experience more inward pull than the object for which boundary is comparatively rough. Thus surface tension is more for a compact structure whereas it is less for a rough structure. More precisely, molecules are loosely bound in a structure for which boundary is rough and thereby this condition facilitates its structural change in terms of size and shape.  It is also a fact that any structural change is a signature of relaxation dynamics in the context of glass forming liquids. Thus, in the present study, we observe a change in $d_{\text{f}}$ with $T_{\text{eff}}$ and $\tau_{p}$ from the insights of morphology variation that causes structural change, both in the BD limit and non-equilibrium limit.
\end{abstract}
\maketitle
\section{\label{intro}Introduction}
The concept of cooperative motion of particles                             \cite{adam1965temperature,blackburn1994cooperative} is often invoked in the context of glassy systems where particles exhibit increasingly sluggish dynamics in  liquids as the  temperature is lowered \cite{ediger1996aca}. This heuristic concept was proposed by Adam and Gibbs \cite{adam1965temperature} in the model of super cooled liquids where relaxation occurs via `cooperatively rearranging regions' (CRRs) that grow in size as temperature is lowered. In the context of passive as well as activity driven glassy system, dynamical heterogeneity and concomitant growing length scales are obvious. Past decades of research in experiments \cite{heuer1995rate, cicerone1995molecules}, simulations \cite{muranaka1995beta, kob1997dynamical} suggest the presence of spatiotemporal heterogeneity where different regions with different mobility relax in different time scales. Elucidating the structure and size of these dynamical heterogeneous regions  give rise the concept of different growing length scales, whereas the morphology of such regions is accompanied by string-like cooperative motion\cite{donati1998stringlike} in the fast subsets for well studied Lennard-Jones (LJ-3D) systems, often treated as a good model for understanding CRRs \cite{kob1997dynamical}. It is also worthy to mention that these regions can also be compact as originally predicted \cite{adam1965temperature,cicerone1997translational}.
The shape of these CRRs also depend on the nature of the interaction potential \cite{sciortino2002one, foffi2005scaling, berthier2009nonperturbative},     \cite{trappe2001jamming,eckert2002re, klix2010structural} between the particles, such as repulsive or attractive potential
\cite{zhang2011cooperative} and control the macroscopic properties of glassy systems \cite{stevenson2006shapes, ediger2000spatially}. In our earlier study\cite{ghoshal2020connecting}, we presented growing dynamical length scale ($\xi$) and relaxation dynamics of an athermal active glass former. Here, in this paper, we describe the structures of   slow and fast moving regions and the notion of fractal dimensions for a synthetic athermal active glass former such as active Ornstein-Uhlenbeck (AOU) type, where persistent propulsion time ($\tau_{p}$) plays an important role  quantitatively as well as qualitatively compared to its passive part.\\
\\
\noindent
In order to discuss in detail, we present our paper in the following sections. Section (\ref{intro}) discusses the introduction and necessary background of the subject. Section (\ref{model}) presents the minimal model and numerical algorithm used in our work to describe an athermal active glass former. Section (\ref{lengthscale}) describes the context and need for cluster substructure analysis. Section (\ref{results}) mentions our results, observations, analysis and finally in Section (\ref{summary}), we  present future directions obtained from our work. Here we describe numerical model as well as algorithm used for cluster substructure analysis.
\section{\label{model} The athermal active Ornstein-Uhlenbeck model.}
The sluggish dynamics of super cooled liquids over observed time scales in the glass transition context is often mentioned as glassy dynamics. This feature is not observed only in the super cooled liquids exhibiting glass transition, but also in some other colloidal systems, complex fluids, driven active matter, biological systems such as bacterial suspension, migrating cells (often mentioned as living active matter) etc. under high dense conditions. The models used for the description of such active matter systems need the inclusion of noise in the governing equations to capture the stochastic nature of such systems. Hence the macroscopic as well as microscopic properties of such systems qualitatively depend on the nature of the noise used in the model to describe such systems.\\
\\
In our analysis we use an athermal Ornstein-Uhlenbeck equation of motion, where particles are self-propelled with respect to the propulsion time and over damped due to high viscosity of the medium. The model is athermal as there is no explicit presence of thermal noise in the position update for the i$^{\text{th}}$ particle of mass $m$ and is described as follows:\\
\begin{eqnarray}
  \bm{\dot r}_{i} &=& \frac{1}{m\gamma}\biggl[- \sum_{j\neq i} \bm \nabla_i \phi_{i} (r_{ij}) + \bm f_i\biggr]\nonumber\\
  \bm{\dot f}_{i} &=& \frac{1}{\tau_p}\biggl[-\bm f_i + \sqrt{2m\gamma k_{\text{B}}T_{\text{eff}}} \boldsymbol{\eta}_{i}\biggr]
\end{eqnarray}
where $m\gamma$ is the friction coefficient in the over damped condition.
Moreover the active Ornstein-Uhlenbeck particles (AOUP) are not kept in an external thermal bath. In case of thermal Ornstein-Uhlenbeck particles, there is always an additional thermal noise along with exponentially correlated colored noise in the governing equation which is absent in our case. Hence, AOUP dynamics is under such a situation where thermal noise for the environment of microswimmers is not considered \cite{caprini2019entropy}. This model stems from the experimental observations where diffusion due to the thermal noise is negligible compared to that due to self-propulsion. This model is applicable for the complex microswimmers such as E.Coli\cite{berg2008coli}, protozoa \cite{blake1974mechanics}, living tissues \cite{poujade2007collective}. Hence there is an additional degree of freedom, namely the self-propulsion. The time correlation of self-propulsion forces $\bm f_i$ follow an exponentially decaying auto correlation function given by\\
    \begin{equation}
      \langle f_{i\alpha}(t)f_{j\beta}(t^{\prime}) \rangle_{\text{noise}} = \biggl(\frac{m \gamma k_{\text{B}}T_{\text{eff}}}{\tau_{p}}\biggr) \delta_{\alpha\beta}\delta_{ij}e^{-|t-t^{\prime}|/\tau_{p}}
   \end{equation}
   where $\alpha$, $\beta$ refer different vector components of this propulsion force and $i$, $j$ are used to denote particle labels. 
  We take  Gaussian white noise $\bm \eta_i$ of zero mean and unit variance and is denoted by
   \begin{equation}
     \langle \bm \eta_{i\alpha}(t) \bm \eta_{j\beta}(t^{\prime}) \rangle_{\text{noise}} = \delta_{\alpha\beta} \delta (t-t') 
     \label{}
   \end{equation}
   where angular brackets $\langle \cdots \rangle$ denote average in the distribution of noise. The particles interact through the Lennard-Jones potential,\\
 \noindent
   \begin{widetext}
   \begin{equation}
     \phi(r_{ij}) = \begin{cases} 
       4\epsilon_{ij} \biggl[ \biggl(\frac{\sigma_{ij}}{r_{ij}}\biggr)^{12} - \biggl(\frac{\sigma_{ij}}{r_{ij}}\biggr)^6\biggr] & \quad 0 < r_{ij} \le r_m \\
       \epsilon_{ij} \biggl[ A \biggl( \frac{\sigma_{ij}}{r_{ij}}\biggr)^{12} -  B\biggl(\frac{\sigma_{ij}}{r_{ij}} \biggr)^6 + \sum_{p=0}^3 C_{2p}\;\biggl(\frac{r_{ij}}{\sigma_{ij}} \biggr)^{2p} \biggr] & \quad r_m < r_{ij} \le r_c\\
       0 & \quad r > r_c
     \end{cases}
     \label{LJ}
   \end{equation}
 \end{widetext}
 where $r_{m}$ and $r_{c}$ are the inner and outer cut-off distances. The values of these distances and other details are mentioned in our earlier work (\cite{ghoshal2020connecting}). Here also we use the same  Kob-Andersen (KA) binary glass former \cite{kob1994scaling, kob1995testing} where the ratio of large($L$) to small ($S$) particles is kept 80:20. We took $\epsilon_{LL}$, $\sigma_{LL}$ and $\sqrt{m \sigma_{LL}^2 / \epsilon_{LL}}$ as the units of energy, length and time, respectively. In these units, the potential parameters become $\epsilon_{SS}$ = 0.5, $\epsilon_{LS}$ = 1.5, $\sigma_{SS}$ = 0.88, and $\sigma_{LS}$ = 0.80.  We have employed a two-dimensional (2D) periodic box having dimension $ 91.287093$ with 10000 particles and density $\rho = 1.2$ that gives us a large enough system size required for cluster substructure analysis associated with growing length scales. We induce the system with moderate range of activity via propulsion time $\tau_p$, ranges from  0.0002 (Brownian dynamics limit or effective equilibrium limit) to a non-equilibrium limit 1.0. We have used stochastic velocity verlet algorithm \cite{mannella1989fast} with a time step $\Delta t=0.0001$ to maintain numerical stability of the simulation for the entire parameter range exploration used in our system. Below we discuss our results.\\
\section{\label{lengthscale} Dynamical length scale and the corresponding cluster substructures}
\begin{figure}[H]
  \centering
\includegraphics[width=0.85\textwidth,keepaspectratio]{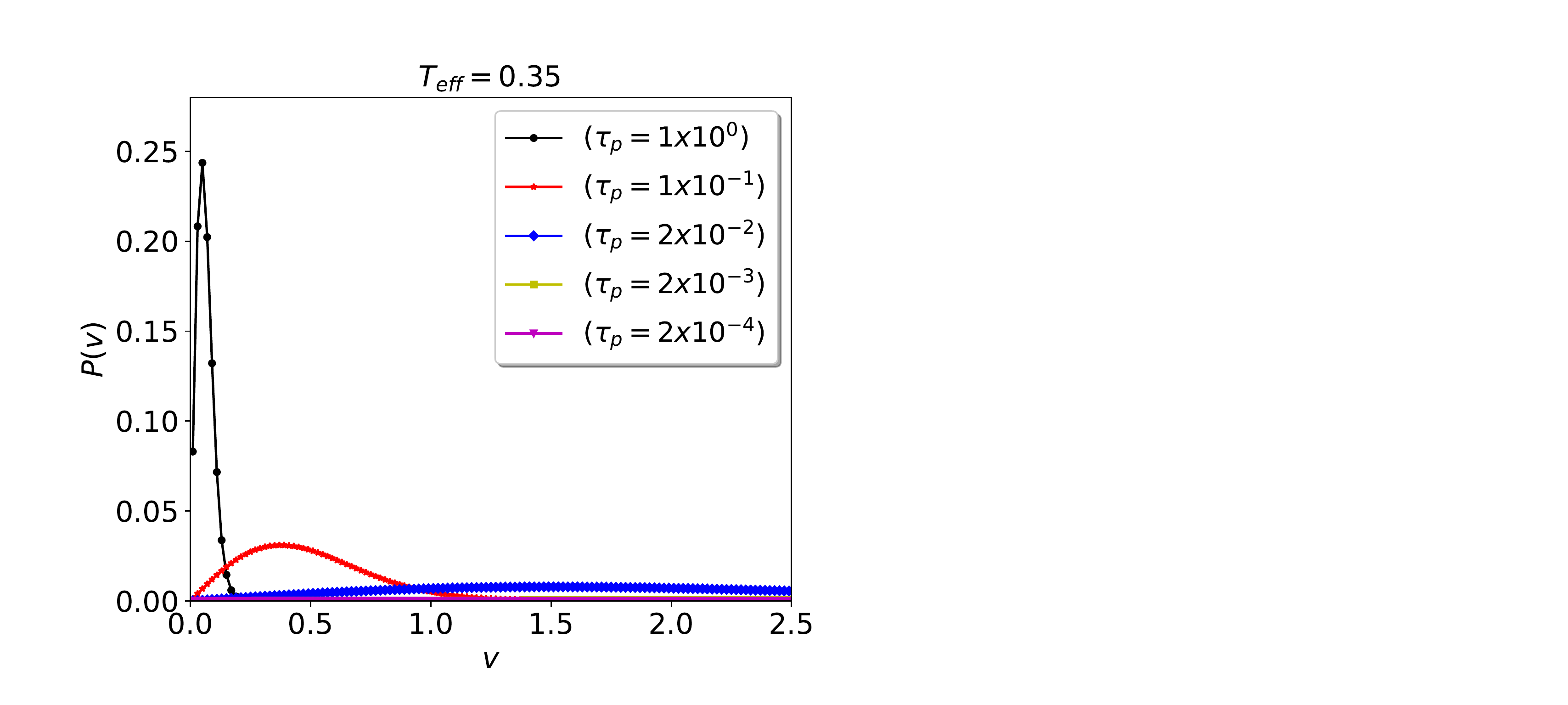}
\caption{Particles become more sluggish as $\tau_{p}$ increases for a fixed $T_{\text{eff}}$. This also confirms that as self-propulsion increases, slow particles are more probable even for lower $T_{\text{eff}}$, such as $0.35$. Thus athermal AOU model enhances glassiness with $\tau_{p}$ rather than diminishing it as opposite to the case predicted by active Brownian particle (ABP) model where strength of the activity fluidify the system.}
\label{vel_dist}
\end{figure}

The growing dynamical length scale for an athermal active  glass former was already extracted in the literature\cite{ghoshal2020connecting}. The effect of $\tau_{p}$ as well as $T_{\text{eff}}$ on this length scale was estimated and it has been seen that size of the clusters obtained from slow moving subsets predicts the size of growing length scale (not shown here) obtained from a four-point structure factor,\\
  \begin{equation}
     S_{4}(q; t) = \frac{1}{N}\biggl(\langle Q(\bm q; t) Q(- \bm q; t)\rangle - |\langle Q(\bm q; t) \rangle|^2  \biggr)
  \end{equation}
    where $Q(\bm q; t)$ is defined as the Fourier transform,
  \begin{equation}
     Q(\bm q; t) = \sum_n w_n(t) \text{exp}\; [-i \bm q \cdot r_n(0)],
  \end{equation}
  of the microscopic overlap function,
    \begin{equation}
  w_{n}(t) = \Theta [a-|\bm r_{n}(t) - \bm r_{n}(0)| ]
  \end{equation}
    Here $\Theta(x)$ is the Heaviside step function and $\bm r_n(t)$ is the position of $n^{\text{th}}$ particle at time $t$. This function helps to filter out the particles that do not move farther than a specified distance $a$, during a time interval $t$ and hence considered as `slow subsets' while the rest of them form `fast subsets'. The value of $a$ corresponds to the plateau value of the mean squared displacement (MSD) at all temperatures (not shown here). For quasi-2D active glassy system, $a=0.2$ in the moderate regime of activity where we observe a plateau like behaviour in the mean-squared displacement of self-propelled particles. We show velocity distribution (Fig. \ref{vel_dist}) for all the particles at a fixed $T_{\text{eff}}=0.35$ and for various $\tau_{p}$. We can see that particles become slow as $\tau_{p}$ increases from BD limit.\\
\noindent
Below we analyze cluster substructures arising from slow and fast subsets for various $T_{\text{eff}}$ and $\tau_{p}$ at different relaxation times. The choice of relaxation times ($\tau_{\alpha}$) are such that the self-intermediate scattering function (SISF) (not shown here) falls $1/e$ of its initial value and where dynamic heterogeneity is maximum.\\
\\
\noindent
It has been mentioned in the literature that formation and relaxation of CRRs occur via fast subsets. But exploration of the same effect is limited in slow subsets. Hence it is our motivation to observe the cluster substructures not only for fast subsets, but for slow subsets as well.
\section{\label{results} Results and Discussion}
\subsection{Cluster substructure analysis through a  density based
spatial clustering algorithm}
\begin{figure}[t]
  \centering
\includegraphics[width=0.85\textwidth,keepaspectratio]{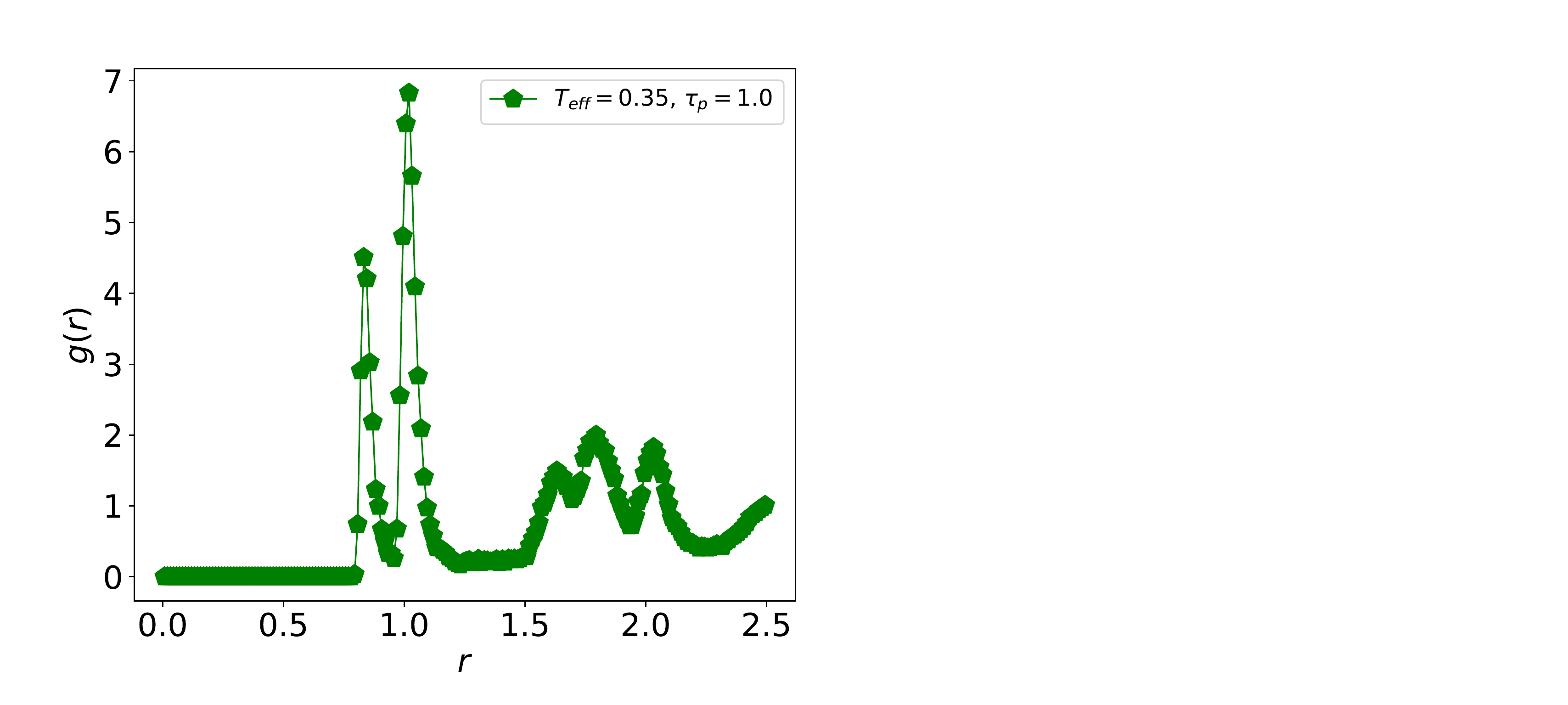}
\caption{The position of first peak for $g(r)$ for all particles comes at around $r\approx 0.83$ for $T_{\text{eff}}=0.35$ and $\tau_{p}=1.0$. The position of the first peak remains the same for all other parameters explored; only peak height differs. Hence we used this position as the epsilon neighbor ($\epsilon$) and considered the same for both fast and slow-subsets.}
\label{g_r}
\end{figure}

\begin{figure}[t]
  \centering
\includegraphics[width=0.85\textwidth,keepaspectratio]{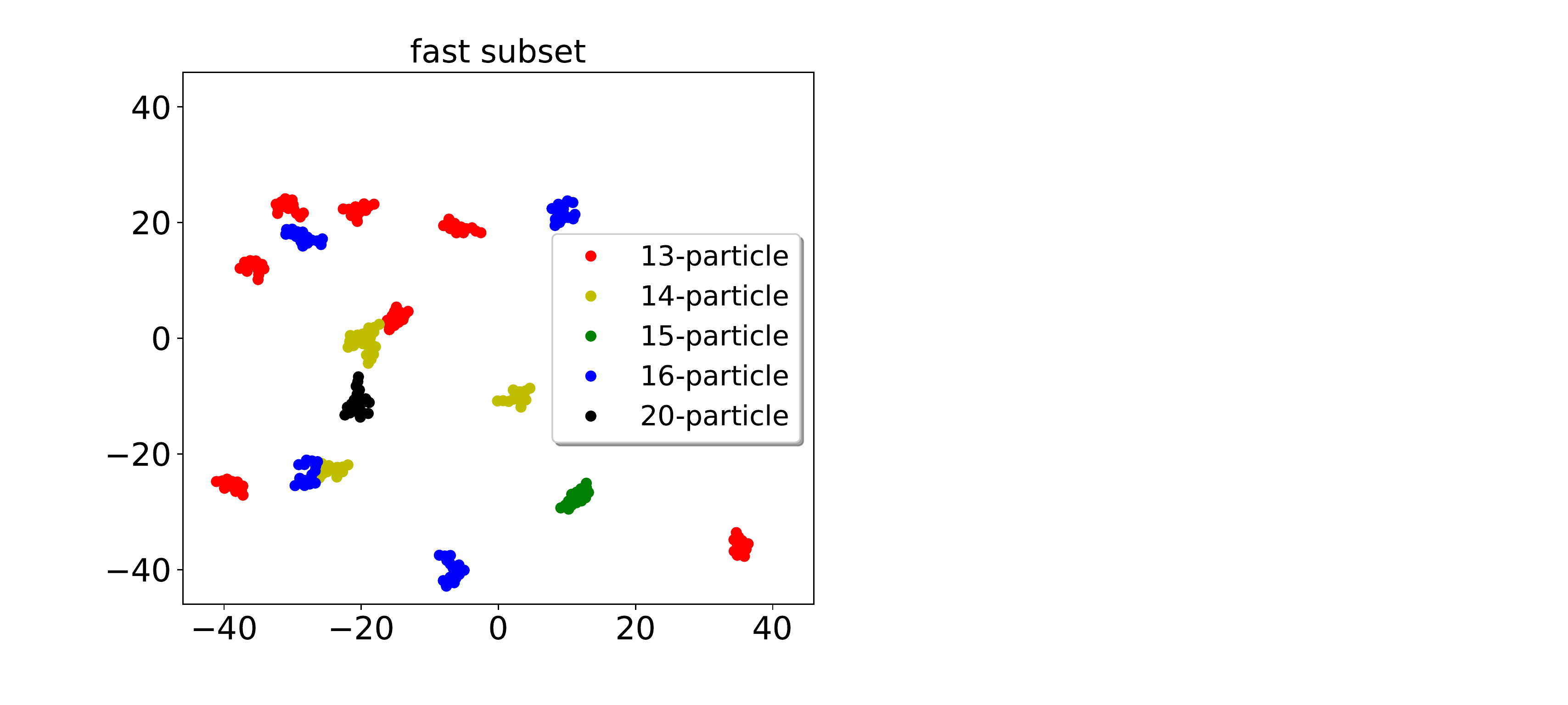}
\includegraphics[width=0.85\textwidth,keepaspectratio]{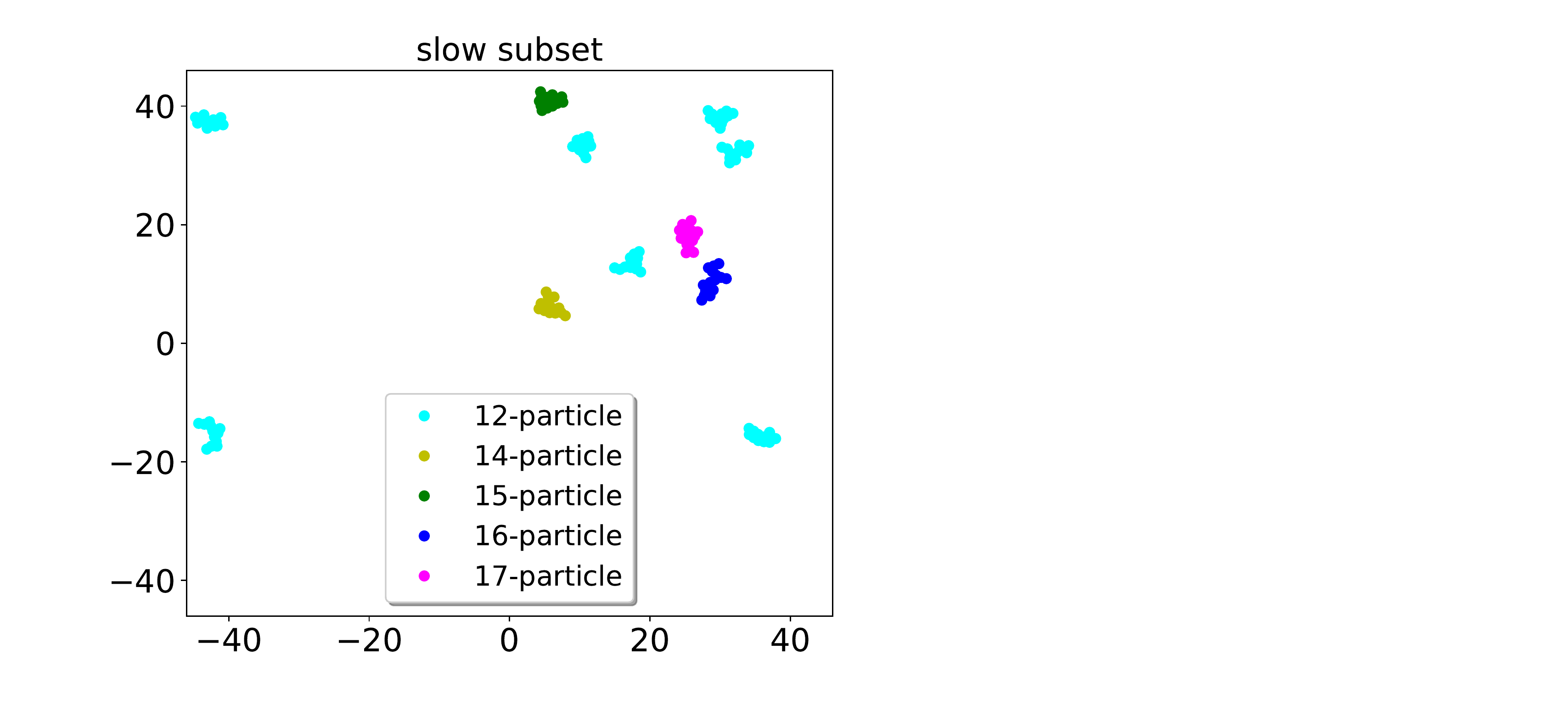}
\caption{This shows few larger particle substructures from fast and slow subsets for $T_{\text{eff}}=0.35$, $\tau_{p}=1.0$. Here various colors represent various particle-structure as mentioned above. There are also smaller particle substructures ranging from 2-particle cluster to 11-particle cluster (not shown here). The frequency for obtaining smaller-particle substructures is more compared to that of larger-particle substructures which is reflected in their probability distribution, mentioned in the next section.}
\label{substructure}
\end{figure}

\begin{figure*}[h]
  \centering
 \includegraphics[width=0.89\textwidth,keepaspectratio]{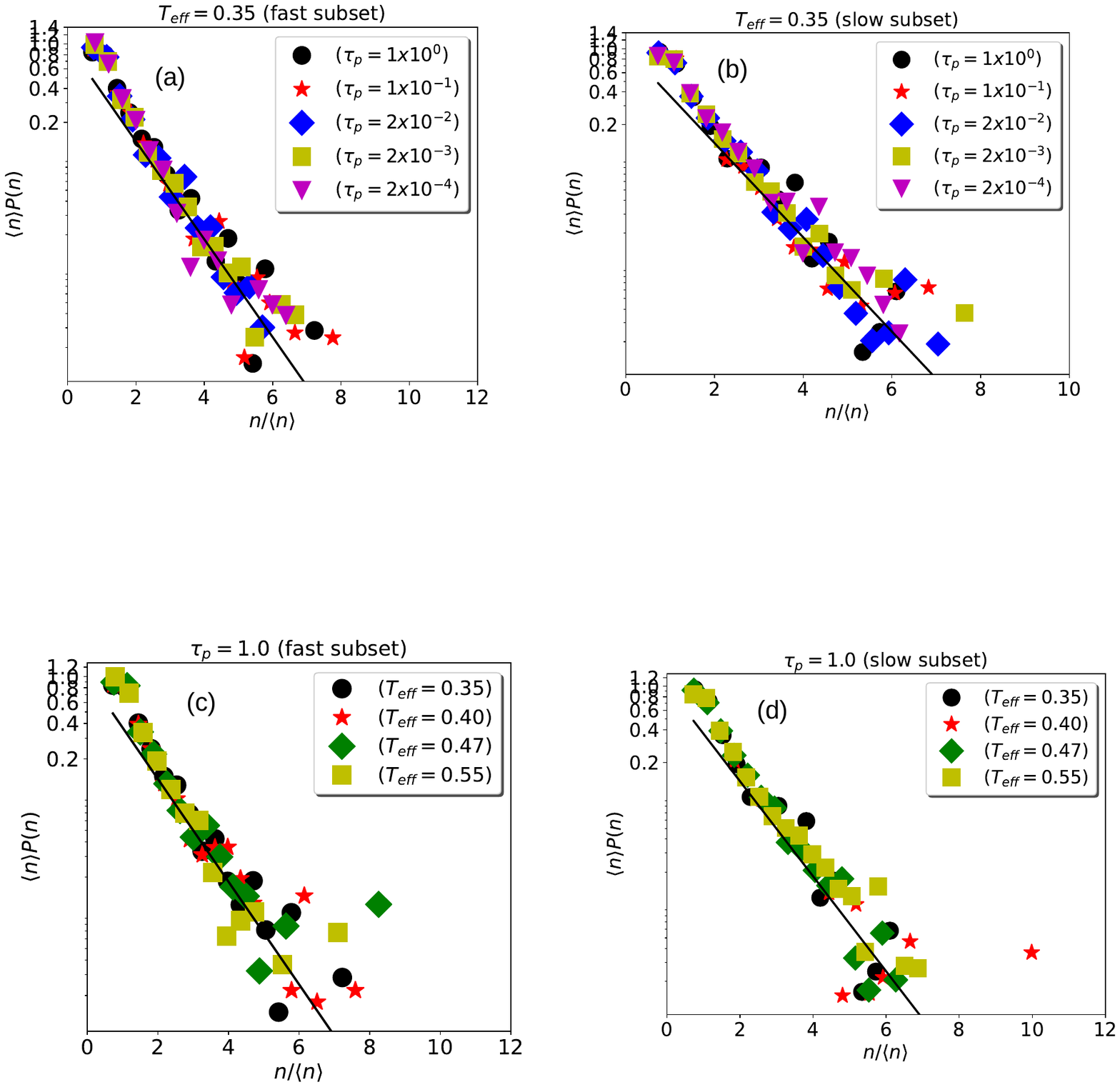} 
\caption{The distribution for substructure length $P(n)$ vs $n$ for fast as well as slow subset  follows as exponential distribution for all  $T_{\text{eff}}$ and $\tau_{p}$. We extract the first moment $\langle n \rangle$ as a fitting parameter of this exponential distribution and do data collapse for both the subsets for a fixed $T_{\text{eff}} = 0.35$ and plotted as $\langle n \rangle P(n)$ vs $n/\langle n \rangle$. The collapsed data also obey the same exponential distribution  as indicated by a single solid line. The average substructure length $\langle n \rangle$ obtained from such distribution also varies in opposite manner (see next Fig(\ref{n_avg})). The value of $\langle n \rangle$ obtained from this distribution is slightly decreasing in magnitude for fast subsets as $T_{\text{eff}}$ increases, whereas the same $\langle n \rangle$ increases as $T_{\text{eff}}$ increases for slow subsets for a fixed $\tau_{p}$ ( see Fig \ref{n_avg}). The same qualitative behaviour is observed for other values of $\tau_{p}$ as $T_{\text{eff}}$ increases in a similar manner for both fast and slow subsets.
}
\label{pn_1}
\end{figure*}

\begin{figure*}[t]
  \centering
\includegraphics[width=0.90\textwidth,keepaspectratio]{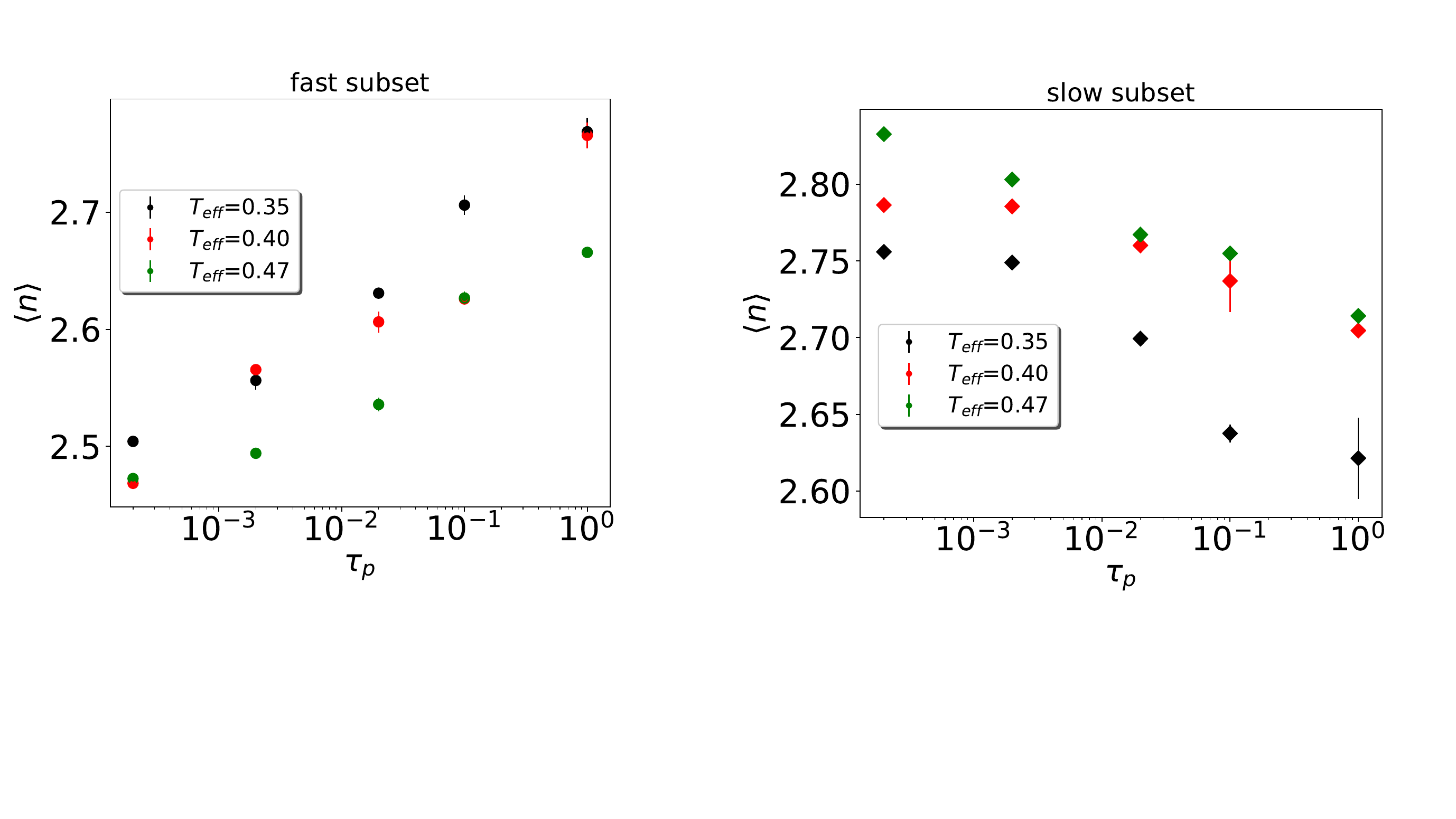}
\caption{$\langle n \rangle$ varies in opposite manner for both fast and slow subsets with respect to $\tau_{p}$ for different $T_{\text{eff}}$. So, the average substructure lengths are distributed in equal and opposite fashion in both the subsets which eventually is reflected in the fractal dimensions. Error bars represent extension of data points around its mean value. Here most of the data points are of the same size as error bar and hence just represented by the size of a single  marker. Thus data for them is concentrated mostly at the mean value, whereas very few data points are scattered around mean value and error bars associated with those data points represent that extension . }
\label{n_avg}
\end{figure*}

There are various clustering algorithms in literature. However, here we use density based spatial clustering algorithm with the application of noise \cite{schubert2017dbscan, sander1998density} to analyze cluster substructures due to their relevance in the present context. According to this protocol, one needs to take a nearest neighbor distance ($\epsilon$-neighbor) for finding neighbors of a given particle. The associated cluster will be formed by all those particles which are $\epsilon$ neighbors to each other. Once a particular cluster is formed for a starting reference particle, often mentioned as `node', we proceed to find next cluster arising from next node. This procedure follows until all the nodes are covered in a given snapshot. The minimum number of points required to construct a cluster is $2$. Hence, we get clusters of different particles, for example, $2$-particle, $3$-particle, $4$-particle, etc. Also, some particles do not participate in cluster formation and remain as `single isolated' particles.\\
\\
Here in glassy dynamics context, we generally take position of the first peak of radial distribution function $g(r)$ as $\epsilon$-neighbour. We see (Fig. (\ref{g_r})) this position is $r\approx 0.83$ for all particles present in the system. The value remains same for all $T_{\text{eff}}$ and $\tau_{p}$ explored in this 2D active glass, only peak height differs for different  activity. In Fig. (\ref{g_r}), we show radial distribution function for full system for a given representative state ($T_{\text{eff}}$, $\tau_{p}$).\\
\noindent
\\
We take those snapshots at a time when the dynamics is mostly heterogeneous. For glass transition perspective, this usually happens at $\alpha$-relaxation time ($\tau_{\alpha}$). Thus we take various snapshots for various initial configurations at $\tau_{\alpha}$ and averaged over all these initial configurations to obtain the final cluster-particle distribution for a given $T_{\text{eff}}$ and $\tau_{p}$ for both fast and slow-subsets.\\
\\
The Fig(\ref{substructure}) shows an example of few particle substructures for a given state $T_{\text{eff}}=0.35$, $\tau_{p}=1.0$. Once we get various particle structures, we see their frequency of occurrence for a given state ($T_{\text{eff}}$, $\tau_{p}$) via histogram (not shown here). Depending on the $\epsilon$ cut-off distance for a given system, small particle structures appear frequently compared to the larger one for each snapshot taken during measurements. We  then plot probability distribution for cluster structure length. Here `length' refers to the number of particles belonging to a particular structure. \\
\\
The structures follow an exponentially decaying function which is reflected in $P(n)$ vs $n$ plot. It reflects the fact that exponential distribution of substructure length  is always associated with cooperative motion  of particles in the dynamics of fast subsets as well as of slow subsets. Thus  the movement of slow subsets also play a role in the relaxation dynamics of CRRs which is not explored in the cooperative movement of passive glass forming particles.\\
\\
Fig(\ref{pn_1}) shows $\langle n \rangle P(n)$ vs $n / \langle n \rangle$ for fast and slow subsets for a fixed $T_{\text{eff}}=0.35$ and various $\tau_{p}$ (Fig \ref{pn_1}(a) and (b)).  Similar trend is also observed for a fixed $\tau_{p} =1.0$ and various $T_{\text{eff}}$ (Fig \ref{pn_1}(c) and (d)).  The $\langle n \rangle$ is the first moment of the distribution and gives the average substructure length. We extract the value of $\langle n \rangle$ from the exponential distribution as a fitting parameter and subsequently show the data collapse. In the next section we mention how one can profitably use $\langle n \rangle$ to extract fractal dimensions ($d_{\text{f}}$) of such structures and can estimate its morphology.\\
\\
The average substructure length $\langle n \rangle$ obtained from the distribution as a fitting parameter tends to vary in opposite manner for fast and slow subsets except with a little bit fluctuation in Brownian dynamics limit, an effective equilibrium regime (see Fig. (\ref{n_avg})). This has the significance towards the shape of a fractal object. Since mass of a fractal object distributed in its various parts depending on the scaling factor and  determines the fractal dimension ($d_{\text{f}}$) of that particular fractal object. So variation of average substructure length with various parameters reflects the variation of its fractal dimensions. Hence we see that effect in $d_{\text{f}}$ for both fast and slow subsets (Fig \ref{df_taup_1}).
It is well known that if mass distribution of a particle aggregate is known, structure and shape of that particle aggregate can also be known from the light scattering experiments and from radius of gyration $R_{g}$ of that particle aggregate. Motivated by this, below we find $R_{g}$ for these particle aggregate to know its shape just by extracting fractal dimensions $d_{\text{f}}$ for both the subsets. Here we mention our findings in a logical order as shown below.
\subsection{Radii of gyration ($R_{g}$) for substructures and associated fractal dimensions ($d_{\text{f}}$)}
In order do find the fractal dimensions $d_{\text{f}}$ for various substructures we need to determine radius of gyration $R_{g}$ of these substructures individually. As $d_{\text{f}}$ follows a power law kind of relationship in the domain of radius of gyration ($R_{g}$) of the small angle scattering intensity plot \cite{lazzari2016fractal} which reveals  structural information of in-homogeneity in large scale system. To calculate $R_{g}$ we use simple equation used in polymer science given as:
\begin{equation}
 R_{g}^{2} = \frac{1}{N} \langle \sum_{i=1}^{N}\left(r_{i} - \bar{r}\right)^{2} \rangle
 \label{radius}
\end{equation}
\begin{figure*}[h]
  \centering
\includegraphics[width=0.91\textwidth,keepaspectratio]{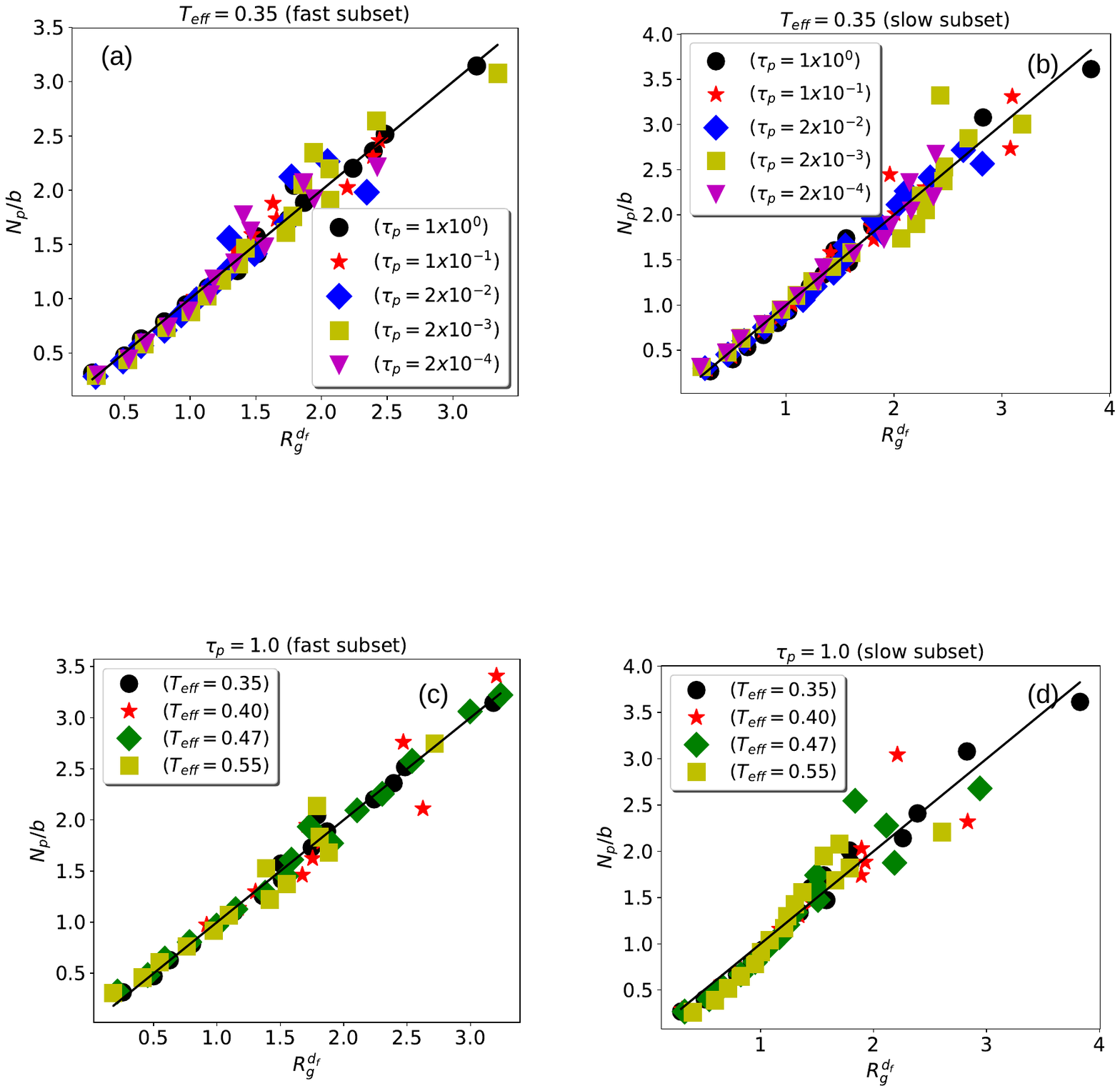}
\caption{Data collapse for the variation of number of particle substructures $N_{p}$ with their $R_{g}$ for a fixed $T_{\text{eff}}=0.35$ for both the subsets and it follows a power law relationship $N_{p}\approx  bR_{g}^{d_{\text{f}}}$, where $b$ has the dimension inverse of $R_{g}$. Fractal dimension $d_{\text{f}}$ for these substructures can readily be obtained from this power law fit. Similar kind of power law variation of $N_{p}$ with their $R_{g}$  is obtained for a fixed $\tau_{p}=1.0$ for both the subsets. We extract $d_{\text{f}}$ from these plots and see its effect on change of pattern for fast and slow subsets.}
\label{r_g_T35}
\end{figure*}
where $\langle\ldots\rangle$ represents ensemble average, $r_{i}$ is the position coordinate of $i^{\text{th}}$ particle comprising the object, $\bar{r}$ represents mean or centre of mass position of the object. $N$ is the total number of particle forming the object.\\
\\
In the present study, we have taken snapshots of particles' positions at a time where dynamics is mostly heterogeneous and separated them as a fast subset and slow subset depending on the distance travel by them (as mentioned in the earlier section (\ref{lengthscale})). Once the position of each particle is known, we group the particles in different substructures like 2-particle, 3-particle and so on depending on the nearest neighbor cut-off $0.83$, obtained from the first peak of $g(r)$. With these information, we calculate $R_{g}$ for each of the substructures separately from equation (\ref{radius}) and plot $N_{p}$ vs $R_{g}$ for each of them at different $T_{\text{eff}}$ and $\tau_{p}$ to see the effect. Clearly, power law relationship $N_{p}\simeq bR_{g}^{d_{\text{f}}}$ (with $d_{\text{f}}$ as an exponent) is obtained for various parameters. The fractal dimensions readily be obtained for each of the control parameters ($T_{\text{eff}},\tau_{p}$) as a fitting parameter for the exponent. We see $d_{\text{f}}$ possesses those values lying in the regime of compact and string like morphology. The $d_{\text{f}}\simeq 1$ corresponds to string-like and $d_{\text{f}}\simeq 2$ for compact like morphology for a fractal object. 
\begin{figure*}[!htbp]
  \centering
\includegraphics[width=0.92\textwidth,keepaspectratio]{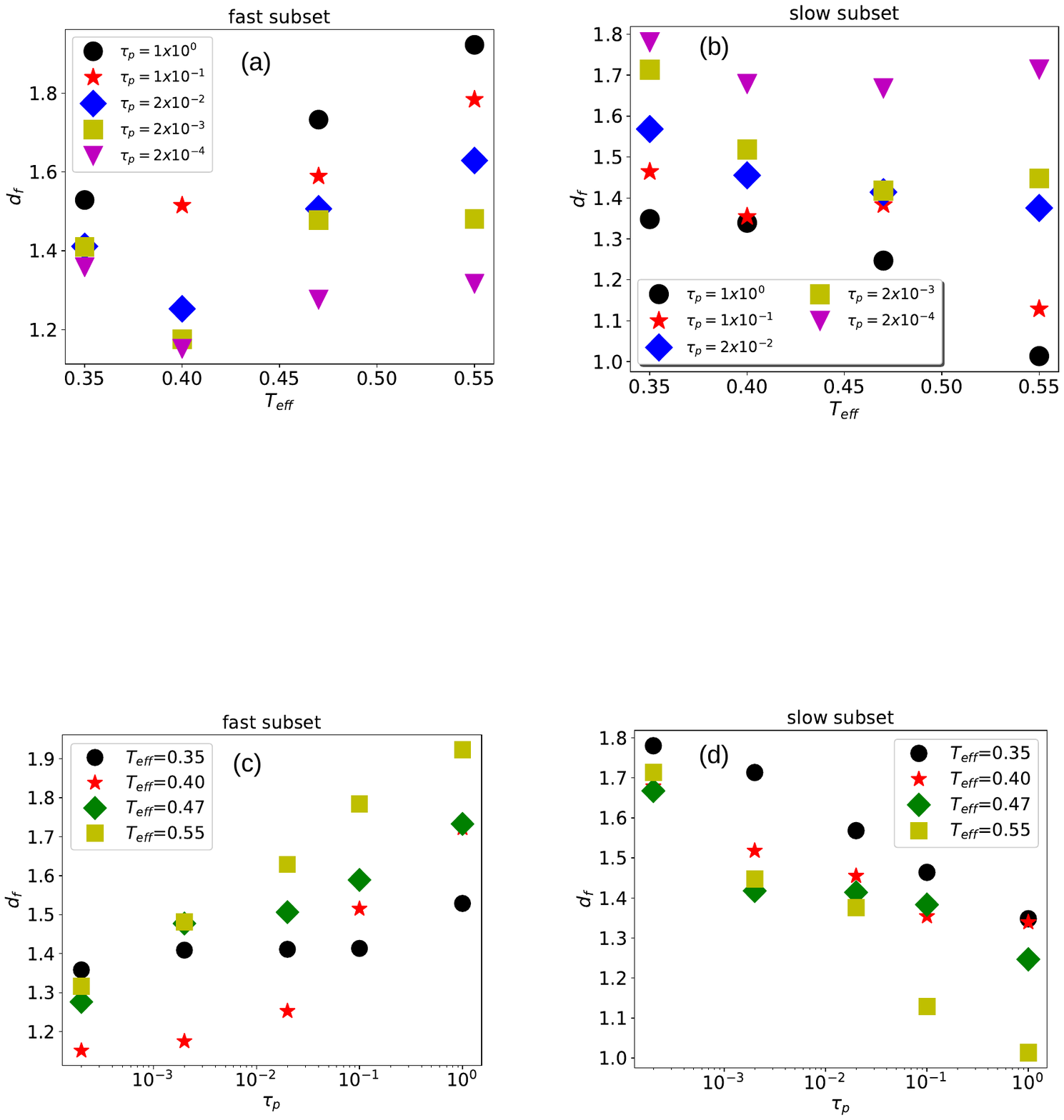}
\caption {Fractal dimensions $d_{\text{f}}$ for all $T_{\text{eff}}$ and for a fixed $\tau_{p}= 1.0$ lies in the range [1.1, 1.8] (Fig. \ref{df_taup_1})(a)), where the lower limit corresponds to string like and upper limit corresponds to compact like morphology. This value for the upper limit holds good for $T_{\text{eff}} = 0.35, 0.40, 0.47$ except at higher $T_{\text{eff}} = 0.55$, where $d_{\text{f}}\cong 2$ (Fig. \ref{df_taup_1})(c)). This observation is at $\tau_{p} = 1$, a far away equilibrium limit for self-propulsion time.  In a similar manner, we can see the opposite effect for slow subsets as explained in the above text.}
\label{df_taup_1}
\end{figure*}
The effect of $\tau_{p}$ (from Fig. (\ref{df_taup_1})(a)) is such that  it changes the morphology of substructures from string to compact for fast moving regions. This change in morphology from string to compact is quite faster for $T_{\text{eff}} = 0.55$ compared to other $T_{\text{eff}}$ explored here and can be easily seen from Fig. (\ref{df_taup_1}(c)). It is also necessary to mention that there is a little bit fluctuation among the $d_{\text{f}}$ values for lower $\tau_{p}$ - the BD limit for different $T_{\text{eff}}$, but this could be of less importance fact as long as $\tau_{p}$ is in the BD limit (Fig. (\ref{df_taup_1}(c))). For higher $\tau_{p}$, change in $d_{\text{f}}$ is prominent systematically leading to the morphology change  from string to compact for fast subsets (Fig. (\ref{df_taup_1})(a) and Fig. (\ref{df_taup_1})(c)). We also observe from Fig. (\ref{df_taup_1})(c) that for lower $T_{\text{eff}}$, such as ($0.35, 0.40$), the values of $d_{\text{f}}$ span only the string like morphology ranging from $1.0$ to $1.7$ irrespective of the $\tau_{p}$ values for fast subsets. But for comparatively higher $T_{\text{eff}}$ ($0.47, 0.55$), the fractal dimension $d_{\text{f}}$ goes from $1.2$ to $2.0$, implying the change of pattern from string to compact as $\tau_{p}$ varies from BD limit to $\tau_{p} = 1$.\\
\\
 For slow subsets, the scenario is slightly different. At lower $T_{\text{eff}}$ regime ($T_{\text{eff}} = 0.35, 0.40$)(Fig. \ref{df_taup_1}(d)), $d_{\text{f}}$ has a tendency to change from string to compact as one goes from higher $\tau_{p} = 1.0$ to lower $\tau_{p}$ (BD limit) value, unlike the fast subset. Similarly for comparatively higher $T_{\text{eff}} = 0.47, 0.55$, variations of $d_{\text{f}}$ with $\tau_{p}$ imply string to compact like morphology as one moves from higher $\tau_{p} = 1.0$ to lower $\tau_{p}$, the BD limit, opposite to that of fast subsets. For example (Fig. \ref{df_taup_1}(d)), at $T_{\text{eff}}=0.47$, the $d_{\text{f}}$ values change from $1.2$ to $1.7$ as one approaches from higher $\tau_{p}=1.0$ to lower $\tau_{p}=2\times10^{-4}$, implying a change from string-like to compact like morphology. The similar trend is also observed for other $T_{\text{eff}}$. Here also a little bit fluctuation is present among the values of $d_{\text{f}}$ at lower $\tau_{p}$ (BD) limit irrespective of various $T_{\text{eff}}$ (Fig. \ref{df_taup_1}(b) and (d)), though the patterns follow same trend for all $T_{\text{eff}}$ and for other higher $\tau_{p}$.\\
\\
Both fast and slow subsets bear string and compact substructures together unlike the past study \cite{donati1998stringlike}, which refers only string-like morphology in fast subset for a passive glassy system.  In the activity induced system, the scenario is quite different.\\
\\
It is well known that the fractal dimension is a measure of the roughness of a fractal object (geometrical structure). So along with the pattern variation, one can predict that roughness is also changing slowly in both fast and slow subsets. In another way, one also can say that $d_{\text{f}}$ gives a measure of the compactness of a fractal object near its boundary. Hence it can imply surface tension at the boundary. For example, if the compactness of an object is low, it will have lower surface tension at the boundary. The object's size can increase quickly because of its less compactness, facilitating structural relaxation. In our analysis, all these estimations are for maximum dynamic heterogeneity that corresponds to $\alpha$ - relaxation time. For a pair of ($T_{\text{eff}}$, $\tau_{p}$), this relaxation time is different. In the previous analysis of our work \cite{ghoshal2020connecting}, we have seen that the effective chemical potential $\mu$ (defined as free energy per length) is only a function of $\tau_{p}$  when free energy for cluster relaxation is scaled with $T_{\text{eff}}$. Besides that size of slowly moving subsets determines the growing dynamical length scale that increases in size for higher $\tau_{p}$ values. This chemical potential $\mu$ decreases as one move towards higher $\tau_{p}$. It suggests that lower values of $\mu$ imply lower surface tension of a fractal object near its boundary, facilitating the generation of larger size clusters as compactness reduces at the boundary. In other words, larger size clusters comprised of those substructures that are more string-like, rendering the increase of roughness and reduction of compactness at the boundaries of a fractal object coming from slow subsets for higher $\tau_{p}$. Here we find exactly similar observation from Fig. (\ref{df_taup_1}(b)) and Fig. (\ref{df_taup_1}(d)) that slow subsets are accompanied by string-like morphology in higher $\tau_{p}$ region irrespective of $T_{\text{eff}}$. The only difference is that here $d_{\text{f}}$ is not solely a function of $\tau_{p}$, but rather a function of both $T_{\text{eff}}$ and $\tau_{p}$. Hence in order to see the dependence of $d_{\text{f}}$ solely with $\tau_{p}$, we need to have an effective dynamics condition. This condition is obtained when fractal dimensions ($d_{\text{f}}$) are measured for the same iso-relaxation time, an effective dynamics condition irrespective of the temperatures ($T_{\text{eff}}$). Then it is possible to obtain $d_{\text{f}}$ variations with respect to $\tau_{p}$ solely and thus one can predict a definite relationship between $\mu$ and $d_{\text{f}}$. This work will be communicated later.
\section{\label{summary} Summary}
The finding of our work is that the exponential distribution of $P(n)$ vs. $n$ is valid for slow subsets and fast subsets, which is a signature of CRRs. Existing consensus towards the morphology of CRRs for passive glassy systems relies only on fast subsets. Our findings can validate the role of slow subsets towards the formation of CRRs regions and their  morphology, not only for passive but also for a synthetic active glass-forming system. We also observe that the fractal patterns vary oppositely for both the subsets and change from string-like morphology to compact-like morphology and vice versa. Another observation is that summation of $d_{\text{f}}$ values for fast and slow subsets are the same throughout the parameter variations and eventually keep the mass distribution the same in the various parts of the fractal object (unless we divide them as fast or slow). We have done extensive simulations for all the parameters, and a detailed cluster substructure analysis is presented here. Since $d_{\text{f}}$ estimates the roughness of a fractal object, in another way, compactness of a fractal structure depends on $d_{\text{f}}$. It hence has implications towards the surface tension of its boundary. Recently the work \cite{li2021softness} by Li \textit{et al.} on the model of cell tissues has revealed the existence of a fractal-like energy landscape. These epithelial cell tissues exhibit slow dynamics, which resembles that of supercooled liquids. So the analysis of our work on cluster substructures and associated fractal dimensions for a synthetic active glass former shed light on the notion of energy landscape indirectly and relaxation dynamics of an athermal active glass former resembling few features of living active matter. Previous studies on macroscopic properties of glasses, such as mechanical, especially yielding \cite{pham2006yielding, zaccarelli2009colloidal, petekidis2004yielding}, rheological properties such as shear thickening \cite{brown2009dynamic, brown2010generality} can also be explored with this concept of change of pattern of fractal substructures and associated fractal dimensions as it measures roughness at the boundary of a fractal object. Therefore, we expect that the present observations will be relevant in 3D active glass and experiments.

\section{Acknowledgements}
We thank Ethayaraja Mani for comments and discussions. All simulations were done on the AQUA super cluster of IIT Madras and the HPC-Physics cluster of our group. Support from DST INSPIRE Faculty Grant 2013/PH-59 and the New Faculty Seed Grant (NFSG), IIT Madras, is gratefully acknowledged.

\bibliographystyle{apsrev}
\bibliography{manuscript}

\begin{thebibliography}{35}
\expandafter\ifx\csname natexlab\endcsname\relax\def\natexlab#1{#1}\fi
\expandafter\ifx\csname bibnamefont\endcsname\relax
  \def\bibnamefont#1{#1}\fi
\expandafter\ifx\csname bibfnamefont\endcsname\relax
  \def\bibfnamefont#1{#1}\fi
\expandafter\ifx\csname citenamefont\endcsname\relax
  \def\citenamefont#1{#1}\fi
\expandafter\ifx\csname url\endcsname\relax
  \def\url#1{\texttt{#1}}\fi
\expandafter\ifx\csname urlprefix\endcsname\relax\def\urlprefix{URL }\fi
\providecommand{\bibinfo}[2]{#2}
\providecommand{\eprint}[2][]{\url{#2}}

\bibitem[{\citenamefont{Adam and Gibbs}(1965)}]{adam1965temperature}
\bibinfo{author}{\bibfnamefont{G.}~\bibnamefont{Adam}} \bibnamefont{and}
  \bibinfo{author}{\bibfnamefont{J.~H.} \bibnamefont{Gibbs}},
  \bibinfo{journal}{The journal of chemical physics}
  \textbf{\bibinfo{volume}{43}}, \bibinfo{pages}{139} (\bibinfo{year}{1965}).

\bibitem[{\citenamefont{Blackburn et~al.}(1994)\citenamefont{Blackburn,
  Cicerone, Hietpas, Wagner, and Ediger}}]{blackburn1994cooperative}
\bibinfo{author}{\bibfnamefont{F.}~\bibnamefont{Blackburn}},
  \bibinfo{author}{\bibfnamefont{M.~T.} \bibnamefont{Cicerone}},
  \bibinfo{author}{\bibfnamefont{G.}~\bibnamefont{Hietpas}},
  \bibinfo{author}{\bibfnamefont{P.~A.} \bibnamefont{Wagner}},
  \bibnamefont{and} \bibinfo{author}{\bibfnamefont{M.}~\bibnamefont{Ediger}},
  \bibinfo{journal}{Journal of non-crystalline solids}
  \textbf{\bibinfo{volume}{172}}, \bibinfo{pages}{256} (\bibinfo{year}{1994}).

\bibitem[{\citenamefont{Ediger}(1996)}]{ediger1996aca}
\bibinfo{author}{\bibfnamefont{M.}~\bibnamefont{Ediger}}, \bibinfo{journal}{J.
  Phys. Chem} \textbf{\bibinfo{volume}{100}}, \bibinfo{pages}{13200}
  (\bibinfo{year}{1996}).

\bibitem[{\citenamefont{Heuer et~al.}(1995)\citenamefont{Heuer, Wilhelm,
  Zimmermann, and Spiess}}]{heuer1995rate}
\bibinfo{author}{\bibfnamefont{A.}~\bibnamefont{Heuer}},
  \bibinfo{author}{\bibfnamefont{M.}~\bibnamefont{Wilhelm}},
  \bibinfo{author}{\bibfnamefont{H.}~\bibnamefont{Zimmermann}},
  \bibnamefont{and} \bibinfo{author}{\bibfnamefont{H.~W.}
  \bibnamefont{Spiess}}, \bibinfo{journal}{Physical review letters}
  \textbf{\bibinfo{volume}{75}}, \bibinfo{pages}{2851} (\bibinfo{year}{1995}).

\bibitem[{\citenamefont{Cicerone et~al.}(1995)\citenamefont{Cicerone,
  Blackburn, and Ediger}}]{cicerone1995molecules}
\bibinfo{author}{\bibfnamefont{M.~T.} \bibnamefont{Cicerone}},
  \bibinfo{author}{\bibfnamefont{F.}~\bibnamefont{Blackburn}},
  \bibnamefont{and} \bibinfo{author}{\bibfnamefont{M.}~\bibnamefont{Ediger}},
  \bibinfo{journal}{The Journal of chemical physics}
  \textbf{\bibinfo{volume}{102}}, \bibinfo{pages}{471} (\bibinfo{year}{1995}).

\bibitem[{\citenamefont{Muranaka and Hiwatari}(1995)}]{muranaka1995beta}
\bibinfo{author}{\bibfnamefont{T.}~\bibnamefont{Muranaka}} \bibnamefont{and}
  \bibinfo{author}{\bibfnamefont{Y.}~\bibnamefont{Hiwatari}},
  \bibinfo{journal}{Physical Review E} \textbf{\bibinfo{volume}{51}},
  \bibinfo{pages}{R2735} (\bibinfo{year}{1995}).

\bibitem[{\citenamefont{Kob et~al.}(1997)\citenamefont{Kob, Donati, Plimpton,
  Poole, and Glotzer}}]{kob1997dynamical}
\bibinfo{author}{\bibfnamefont{W.}~\bibnamefont{Kob}},
  \bibinfo{author}{\bibfnamefont{C.}~\bibnamefont{Donati}},
  \bibinfo{author}{\bibfnamefont{S.~J.} \bibnamefont{Plimpton}},
  \bibinfo{author}{\bibfnamefont{P.~H.} \bibnamefont{Poole}}, \bibnamefont{and}
  \bibinfo{author}{\bibfnamefont{S.~C.} \bibnamefont{Glotzer}},
  \bibinfo{journal}{Physical review letters} \textbf{\bibinfo{volume}{79}},
  \bibinfo{pages}{2827} (\bibinfo{year}{1997}).

\bibitem[{\citenamefont{Donati et~al.}(1998)\citenamefont{Donati, Douglas, Kob,
  Plimpton, Poole, and Glotzer}}]{donati1998stringlike}
\bibinfo{author}{\bibfnamefont{C.}~\bibnamefont{Donati}},
  \bibinfo{author}{\bibfnamefont{J.~F.} \bibnamefont{Douglas}},
  \bibinfo{author}{\bibfnamefont{W.}~\bibnamefont{Kob}},
  \bibinfo{author}{\bibfnamefont{S.~J.} \bibnamefont{Plimpton}},
  \bibinfo{author}{\bibfnamefont{P.~H.} \bibnamefont{Poole}}, \bibnamefont{and}
  \bibinfo{author}{\bibfnamefont{S.~C.} \bibnamefont{Glotzer}},
  \bibinfo{journal}{Physical review letters} \textbf{\bibinfo{volume}{80}},
  \bibinfo{pages}{2338} (\bibinfo{year}{1998}).

\bibitem[{\citenamefont{Cicerone et~al.}(1997)\citenamefont{Cicerone, Wagner,
  and Ediger}}]{cicerone1997translational}
\bibinfo{author}{\bibfnamefont{M.~T.} \bibnamefont{Cicerone}},
  \bibinfo{author}{\bibfnamefont{P.~A.} \bibnamefont{Wagner}},
  \bibnamefont{and} \bibinfo{author}{\bibfnamefont{M.}~\bibnamefont{Ediger}},
  \bibinfo{journal}{The Journal of Physical Chemistry B}
  \textbf{\bibinfo{volume}{101}}, \bibinfo{pages}{8727} (\bibinfo{year}{1997}).

\bibitem[{\citenamefont{Sciortino}(2002)}]{sciortino2002one}
\bibinfo{author}{\bibfnamefont{F.}~\bibnamefont{Sciortino}},
  \bibinfo{journal}{Nature materials} \textbf{\bibinfo{volume}{1}},
  \bibinfo{pages}{145} (\bibinfo{year}{2002}).

\bibitem[{\citenamefont{Foffi et~al.}(2005)\citenamefont{Foffi, De~Michele,
  Sciortino, and Tartaglia}}]{foffi2005scaling}
\bibinfo{author}{\bibfnamefont{G.}~\bibnamefont{Foffi}},
  \bibinfo{author}{\bibfnamefont{C.}~\bibnamefont{De~Michele}},
  \bibinfo{author}{\bibfnamefont{F.}~\bibnamefont{Sciortino}},
  \bibnamefont{and}
  \bibinfo{author}{\bibfnamefont{P.}~\bibnamefont{Tartaglia}},
  \bibinfo{journal}{Physical review letters} \textbf{\bibinfo{volume}{94}},
  \bibinfo{pages}{078301} (\bibinfo{year}{2005}).

\bibitem[{\citenamefont{Berthier and
  Tarjus}(2009)}]{berthier2009nonperturbative}
\bibinfo{author}{\bibfnamefont{L.}~\bibnamefont{Berthier}} \bibnamefont{and}
  \bibinfo{author}{\bibfnamefont{G.}~\bibnamefont{Tarjus}},
  \bibinfo{journal}{Physical review letters} \textbf{\bibinfo{volume}{103}},
  \bibinfo{pages}{170601} (\bibinfo{year}{2009}).

\bibitem[{\citenamefont{Trappe et~al.}(2001)\citenamefont{Trappe, Prasad,
  Cipelletti, Segre, and Weitz}}]{trappe2001jamming}
\bibinfo{author}{\bibfnamefont{V.}~\bibnamefont{Trappe}},
  \bibinfo{author}{\bibfnamefont{V.}~\bibnamefont{Prasad}},
  \bibinfo{author}{\bibfnamefont{L.}~\bibnamefont{Cipelletti}},
  \bibinfo{author}{\bibfnamefont{P.}~\bibnamefont{Segre}}, \bibnamefont{and}
  \bibinfo{author}{\bibfnamefont{D.~A.} \bibnamefont{Weitz}},
  \bibinfo{journal}{Nature} \textbf{\bibinfo{volume}{411}},
  \bibinfo{pages}{772} (\bibinfo{year}{2001}).

\bibitem[{\citenamefont{Eckert and Bartsch}(2002)}]{eckert2002re}
\bibinfo{author}{\bibfnamefont{T.}~\bibnamefont{Eckert}} \bibnamefont{and}
  \bibinfo{author}{\bibfnamefont{E.}~\bibnamefont{Bartsch}},
  \bibinfo{journal}{Physical review letters} \textbf{\bibinfo{volume}{89}},
  \bibinfo{pages}{125701} (\bibinfo{year}{2002}).

\bibitem[{\citenamefont{Klix et~al.}(2010)\citenamefont{Klix, Royall, and
  Tanaka}}]{klix2010structural}
\bibinfo{author}{\bibfnamefont{C.~L.} \bibnamefont{Klix}},
  \bibinfo{author}{\bibfnamefont{C.~P.} \bibnamefont{Royall}},
  \bibnamefont{and} \bibinfo{author}{\bibfnamefont{H.}~\bibnamefont{Tanaka}},
  \bibinfo{journal}{Physical review letters} \textbf{\bibinfo{volume}{104}},
  \bibinfo{pages}{165702} (\bibinfo{year}{2010}).

\bibitem[{\citenamefont{Zhang et~al.}(2011)\citenamefont{Zhang, Yunker, Habdas,
  and Yodh}}]{zhang2011cooperative}
\bibinfo{author}{\bibfnamefont{Z.}~\bibnamefont{Zhang}},
  \bibinfo{author}{\bibfnamefont{P.~J.} \bibnamefont{Yunker}},
  \bibinfo{author}{\bibfnamefont{P.}~\bibnamefont{Habdas}}, \bibnamefont{and}
  \bibinfo{author}{\bibfnamefont{A.}~\bibnamefont{Yodh}},
  \bibinfo{journal}{Physical review letters} \textbf{\bibinfo{volume}{107}},
  \bibinfo{pages}{208303} (\bibinfo{year}{2011}).

\bibitem[{\citenamefont{Stevenson et~al.}(2006)\citenamefont{Stevenson,
  Schmalian, and Wolynes}}]{stevenson2006shapes}
\bibinfo{author}{\bibfnamefont{J.~D.} \bibnamefont{Stevenson}},
  \bibinfo{author}{\bibfnamefont{J.}~\bibnamefont{Schmalian}},
  \bibnamefont{and} \bibinfo{author}{\bibfnamefont{P.~G.}
  \bibnamefont{Wolynes}}, \bibinfo{journal}{Nature Physics}
  \textbf{\bibinfo{volume}{2}}, \bibinfo{pages}{268} (\bibinfo{year}{2006}).

\bibitem[{\citenamefont{Ediger}(2000)}]{ediger2000spatially}
\bibinfo{author}{\bibfnamefont{M.~D.} \bibnamefont{Ediger}},
  \bibinfo{journal}{Annual review of physical chemistry}
  \textbf{\bibinfo{volume}{51}}, \bibinfo{pages}{99} (\bibinfo{year}{2000}).

\bibitem[{\citenamefont{Ghoshal and Joy}(2020)}]{ghoshal2020connecting}
\bibinfo{author}{\bibfnamefont{D.}~\bibnamefont{Ghoshal}} \bibnamefont{and}
  \bibinfo{author}{\bibfnamefont{A.}~\bibnamefont{Joy}},
  \bibinfo{journal}{Physical Review E} \textbf{\bibinfo{volume}{102}},
  \bibinfo{pages}{062605} (\bibinfo{year}{2020}).

\bibitem[{\citenamefont{Caprini et~al.}(2019)\citenamefont{Caprini, Marconi,
  Puglisi, and Vulpiani}}]{caprini2019entropy}
\bibinfo{author}{\bibfnamefont{L.}~\bibnamefont{Caprini}},
  \bibinfo{author}{\bibfnamefont{U.~M.~B.} \bibnamefont{Marconi}},
  \bibinfo{author}{\bibfnamefont{A.}~\bibnamefont{Puglisi}}, \bibnamefont{and}
  \bibinfo{author}{\bibfnamefont{A.}~\bibnamefont{Vulpiani}},
  \bibinfo{journal}{Journal of Statistical Mechanics: Theory and Experiment}
  \textbf{\bibinfo{volume}{2019}}, \bibinfo{pages}{053203}
  (\bibinfo{year}{2019}).

\bibitem[{\citenamefont{Berg}(2008)}]{berg2008coli}
\bibinfo{author}{\bibfnamefont{H.~C.} \bibnamefont{Berg}},
  \emph{\bibinfo{title}{E. coli in Motion}} (\bibinfo{publisher}{Springer
  Science \& Business Media}, \bibinfo{year}{2008}).

\bibitem[{\citenamefont{Blake and Sleigh}(1974)}]{blake1974mechanics}
\bibinfo{author}{\bibfnamefont{J.~R.} \bibnamefont{Blake}} \bibnamefont{and}
  \bibinfo{author}{\bibfnamefont{M.~A.} \bibnamefont{Sleigh}},
  \bibinfo{journal}{Biological Reviews} \textbf{\bibinfo{volume}{49}},
  \bibinfo{pages}{85} (\bibinfo{year}{1974}).

\bibitem[{\citenamefont{Poujade et~al.}(2007)\citenamefont{Poujade,
  Grasland-Mongrain, Hertzog, Jouanneau, Chavrier, Ladoux, Buguin, and
  Silberzan}}]{poujade2007collective}
\bibinfo{author}{\bibfnamefont{M.}~\bibnamefont{Poujade}},
  \bibinfo{author}{\bibfnamefont{E.}~\bibnamefont{Grasland-Mongrain}},
  \bibinfo{author}{\bibfnamefont{A.}~\bibnamefont{Hertzog}},
  \bibinfo{author}{\bibfnamefont{J.}~\bibnamefont{Jouanneau}},
  \bibinfo{author}{\bibfnamefont{P.}~\bibnamefont{Chavrier}},
  \bibinfo{author}{\bibfnamefont{B.}~\bibnamefont{Ladoux}},
  \bibinfo{author}{\bibfnamefont{A.}~\bibnamefont{Buguin}}, \bibnamefont{and}
  \bibinfo{author}{\bibfnamefont{P.}~\bibnamefont{Silberzan}},
  \bibinfo{journal}{Proceedings of the National Academy of Sciences}
  \textbf{\bibinfo{volume}{104}}, \bibinfo{pages}{15988}
  (\bibinfo{year}{2007}).

\bibitem[{\citenamefont{Kob and Andersen}(1994)}]{kob1994scaling}
\bibinfo{author}{\bibfnamefont{W.}~\bibnamefont{Kob}} \bibnamefont{and}
  \bibinfo{author}{\bibfnamefont{H.~C.} \bibnamefont{Andersen}},
  \bibinfo{journal}{Physical review letters} \textbf{\bibinfo{volume}{73}},
  \bibinfo{pages}{1376} (\bibinfo{year}{1994}).

\bibitem[{\citenamefont{Kob and Andersen}(1995)}]{kob1995testing}
\bibinfo{author}{\bibfnamefont{W.}~\bibnamefont{Kob}} \bibnamefont{and}
  \bibinfo{author}{\bibfnamefont{H.~C.} \bibnamefont{Andersen}},
  \bibinfo{journal}{Physical Review E} \textbf{\bibinfo{volume}{51}},
  \bibinfo{pages}{4626} (\bibinfo{year}{1995}).

\bibitem[{\citenamefont{Mannella and Palleschi}(1989)}]{mannella1989fast}
\bibinfo{author}{\bibfnamefont{R.}~\bibnamefont{Mannella}} \bibnamefont{and}
  \bibinfo{author}{\bibfnamefont{V.}~\bibnamefont{Palleschi}},
  \bibinfo{journal}{Physical Review A} \textbf{\bibinfo{volume}{40}},
  \bibinfo{pages}{3381} (\bibinfo{year}{1989}).

\bibitem[{\citenamefont{Schubert et~al.}(2017)\citenamefont{Schubert, Sander,
  Ester, Kriegel, and Xu}}]{schubert2017dbscan}
\bibinfo{author}{\bibfnamefont{E.}~\bibnamefont{Schubert}},
  \bibinfo{author}{\bibfnamefont{J.}~\bibnamefont{Sander}},
  \bibinfo{author}{\bibfnamefont{M.}~\bibnamefont{Ester}},
  \bibinfo{author}{\bibfnamefont{H.~P.} \bibnamefont{Kriegel}},
  \bibnamefont{and} \bibinfo{author}{\bibfnamefont{X.}~\bibnamefont{Xu}},
  \bibinfo{journal}{ACM Transactions on Database Systems (TODS)}
  \textbf{\bibinfo{volume}{42}}, \bibinfo{pages}{1} (\bibinfo{year}{2017}).

\bibitem[{\citenamefont{Sander et~al.}(1998)\citenamefont{Sander, Ester,
  Kriegel, and Xu}}]{sander1998density}
\bibinfo{author}{\bibfnamefont{J.}~\bibnamefont{Sander}},
  \bibinfo{author}{\bibfnamefont{M.}~\bibnamefont{Ester}},
  \bibinfo{author}{\bibfnamefont{H.-P.} \bibnamefont{Kriegel}},
  \bibnamefont{and} \bibinfo{author}{\bibfnamefont{X.}~\bibnamefont{Xu}},
  \bibinfo{journal}{Data mining and knowledge discovery}
  \textbf{\bibinfo{volume}{2}}, \bibinfo{pages}{169} (\bibinfo{year}{1998}).

\bibitem[{\citenamefont{Lazzari et~al.}(2016)\citenamefont{Lazzari, Nicoud,
  Jaquet, Lattuada, and Morbidelli}}]{lazzari2016fractal}
\bibinfo{author}{\bibfnamefont{S.}~\bibnamefont{Lazzari}},
  \bibinfo{author}{\bibfnamefont{L.}~\bibnamefont{Nicoud}},
  \bibinfo{author}{\bibfnamefont{B.}~\bibnamefont{Jaquet}},
  \bibinfo{author}{\bibfnamefont{M.}~\bibnamefont{Lattuada}}, \bibnamefont{and}
  \bibinfo{author}{\bibfnamefont{M.}~\bibnamefont{Morbidelli}},
  \bibinfo{journal}{Advances in colloid and interface science}
  \textbf{\bibinfo{volume}{235}}, \bibinfo{pages}{1} (\bibinfo{year}{2016}).

\bibitem[{\citenamefont{Li et~al.}(2021)\citenamefont{Li, Wei, Paoluzzi, and
  Ciamarra}}]{li2021softness}
\bibinfo{author}{\bibfnamefont{Y.-W.} \bibnamefont{Li}},
  \bibinfo{author}{\bibfnamefont{L.~L.~Y.} \bibnamefont{Wei}},
  \bibinfo{author}{\bibfnamefont{M.}~\bibnamefont{Paoluzzi}}, \bibnamefont{and}
  \bibinfo{author}{\bibfnamefont{M.~P.} \bibnamefont{Ciamarra}},
  \bibinfo{journal}{Physical Review E} \textbf{\bibinfo{volume}{103}},
  \bibinfo{pages}{022607} (\bibinfo{year}{2021}).

\bibitem[{\citenamefont{Pham et~al.}(2006)\citenamefont{Pham, Petekidis,
  Vlassopoulos, Egelhaaf, Pusey, and Poon}}]{pham2006yielding}
\bibinfo{author}{\bibfnamefont{K.}~\bibnamefont{Pham}},
  \bibinfo{author}{\bibfnamefont{G.}~\bibnamefont{Petekidis}},
  \bibinfo{author}{\bibfnamefont{D.}~\bibnamefont{Vlassopoulos}},
  \bibinfo{author}{\bibfnamefont{S.}~\bibnamefont{Egelhaaf}},
  \bibinfo{author}{\bibfnamefont{P.}~\bibnamefont{Pusey}}, \bibnamefont{and}
  \bibinfo{author}{\bibfnamefont{W.}~\bibnamefont{Poon}}, \bibinfo{journal}{EPL
  (Europhysics Letters)} \textbf{\bibinfo{volume}{75}}, \bibinfo{pages}{624}
  (\bibinfo{year}{2006}).

\bibitem[{\citenamefont{Zaccarelli and Poon}(2009)}]{zaccarelli2009colloidal}
\bibinfo{author}{\bibfnamefont{E.}~\bibnamefont{Zaccarelli}} \bibnamefont{and}
  \bibinfo{author}{\bibfnamefont{W.~C.} \bibnamefont{Poon}},
  \bibinfo{journal}{Proceedings of the National Academy of Sciences}
  \textbf{\bibinfo{volume}{106}}, \bibinfo{pages}{15203}
  (\bibinfo{year}{2009}).

\bibitem[{\citenamefont{Petekidis et~al.}(2004)\citenamefont{Petekidis,
  Vlassopoulos, and Pusey}}]{petekidis2004yielding}
\bibinfo{author}{\bibfnamefont{G.}~\bibnamefont{Petekidis}},
  \bibinfo{author}{\bibfnamefont{D.}~\bibnamefont{Vlassopoulos}},
  \bibnamefont{and} \bibinfo{author}{\bibfnamefont{P.}~\bibnamefont{Pusey}},
  \bibinfo{journal}{Journal of physics: Condensed matter}
  \textbf{\bibinfo{volume}{16}}, \bibinfo{pages}{S3955} (\bibinfo{year}{2004}).

\bibitem[{\citenamefont{Brown and Jaeger}(2009)}]{brown2009dynamic}
\bibinfo{author}{\bibfnamefont{E.}~\bibnamefont{Brown}} \bibnamefont{and}
  \bibinfo{author}{\bibfnamefont{H.~M.} \bibnamefont{Jaeger}},
  \bibinfo{journal}{Physical review letters} \textbf{\bibinfo{volume}{103}},
  \bibinfo{pages}{086001} (\bibinfo{year}{2009}).

\bibitem[{\citenamefont{Brown et~al.}(2010)\citenamefont{Brown, Forman,
  Orellana, Zhang, Maynor, Betts, DeSimone, and Jaeger}}]{brown2010generality}
\bibinfo{author}{\bibfnamefont{E.}~\bibnamefont{Brown}},
  \bibinfo{author}{\bibfnamefont{N.~A.} \bibnamefont{Forman}},
  \bibinfo{author}{\bibfnamefont{C.~S.} \bibnamefont{Orellana}},
  \bibinfo{author}{\bibfnamefont{H.}~\bibnamefont{Zhang}},
  \bibinfo{author}{\bibfnamefont{B.~W.} \bibnamefont{Maynor}},
  \bibinfo{author}{\bibfnamefont{D.~E.} \bibnamefont{Betts}},
  \bibinfo{author}{\bibfnamefont{J.~M.} \bibnamefont{DeSimone}},
  \bibnamefont{and} \bibinfo{author}{\bibfnamefont{H.~M.}
  \bibnamefont{Jaeger}}, \bibinfo{journal}{Nature materials}
  \textbf{\bibinfo{volume}{9}}, \bibinfo{pages}{220} (\bibinfo{year}{2010}).

\end{thebibliography}
\end{document}